\documentclass{iaus}

\usepackage{graphicx}

\newcommand{\beq}	{\begin{equation}}
\newcommand{\eeq}	{\end{equation}}
\newcommand{\beqa}{\begin{eqnarray}}
\newcommand{\eeqa}{\end{eqnarray}}
\newcommand{\avg}[1]  {{\langle #1 \rangle}} 

\newcommand{\e}	{$^{-1}$}

\def\simlt{\lower.5ex\hbox{$\; \buildrel < \over \sim \;$}}
\def\simgt{\lower.5ex\hbox{$\; \buildrel > \over \sim \;$}}
\def\la{\simlt}

\font\tenbi=cmmib10 
\newfam\bifam  \textfont\bifam=\tenbi

\font\tenbr=cmbx10
\newfam\brfam  \textfont\brfam=\tenbr

\font\squinttenbi=cmbx10 at 9pt
\scriptfont\brfam=\squinttenbi

\def\vecnabla{
            \setbox1=\hbox{$\bigtriangledown$}
                         \raise.45ex\hbox{$\bigtriangledown$\hskip-.97\wd1
                         $\bigtriangledown$\hskip-.97\wd1
                         $\bigtriangledown$\hskip-.97\wd1}
                         \raise.47ex\hbox{$\bigtriangledown$}}

\newcommand{\caln}		{{\cal N}}

\newcommand{\acc}	{{\rm acc}}
\newcommand{\high}	{{\rm high}}
\newcommand{\low}	{{\rm low}}
\newcommand{\dth}	{\Delta t_\high}

\newcommand{\mdh}	{{\dot m_\high}}
\newcommand{\mf} 	{{m_f}}
\newcommand{\ml}		{{m_\ell}}
\newcommand{\mup}		{{m_u}}
\newcommand{\ppt}		{\psi_{p2}}
\newcommand{\Ppl}		{\Psi_p(L)}
\newcommand{\Ppt}		{\Psi_{p2}}
\newcommand{\tf} 	{{t_f}}

\begin{document}
\title[The Luminosity Problem]{The Luminosity Problem: Testing Theories of Star Formation}

\author[C. F. McKee \& S. S. R. Offner]{Christopher F. McKee$^1$ \&
Stella S. R. Offner$^2$
}

\affiliation{$^1$Departments of Physics and Astronomy, University of California, Berkeley, CA94720, USA and\\ Laboratoire d'Etudes du Rayonnement et de la Mati\`ere en Astrophysique, LERMA-LRA, Ecole Normale Superieure, 24 rue Lhomond, 75005 Paris, 
France \\email:{\tt cmckee@astro.berkeley.edu}
\\ $^2$Harvard-Smithsonian Center for Astrophysics, 60 Garden St,
Cambridge MA 02138, USA}

\pubyear{2010}
\volume{270}  
\pagerange{}
\setcounter{page}{1}
\jname{Computational Star Formation}
\editors{A.C. Editor, B.D. Editor \& C.E. Editor, eds.}

\maketitle

\begin{abstract}
Low-mass protostars are less luminous than expected. This luminosity problem
is important because the observations appear to be inconsistent with 
some of the basic premises of star formation theory. Two possible solutions
are that stars form 
slowly, which is supported
by recent data, and/or that protostellar accretion is episodic; current 
data suggest that the latter
accounts for less than half the missing luminosity. 
The solution to the luminosity problem bears directly on the fundamental problem
of the time required to form a low-mass star.
The protostellar mass and
luminosity functions provide powerful tools both for addressing the luminosity
problem and for testing theories of star formation. 
Results are presented for the 
collapse of singular isothermal spheres, for the collapse of turbulent cores, and
for competitive accretion.

\end{abstract}

\section{The Luminosity Problem}

Why don't protostars shine more brightly? In a seminal paper,
Kenyon et al. (1990) identified this 
luminosity problem,
developed an approach to treat it (the protostellar luminosity function),
and proposed almost all the solutions to the problem that have subsequently been
studied. The luminosity problem is simple to
state: The accretion luminosity of a protostar is
\beq
L_\acc=f_\acc\; \frac{Gm \dot m}{r_*}=
3.9 f_\acc\left(\frac{m}{0.25 M_\odot}\right)\left(\frac{2 R_\odot}{r_*}\right)
\left(\frac{\dot m}{10^{-6}\,M_\odot\, \mbox{yr\e}}\right)~~~L_\odot,
\label{eq:lacc}
\eeq
where $f_\acc$ is the fraction of the accretion energy that goes into radiation
(Kenyon et al. took $f_\acc=1$),
$m$ is the protostellar mass, and $\dot m$ is the accretion rate.
Stahler (1988) has calculated the protostellar radius for a variety of
masses and accretion rates, including the thermostatic effects of deuterium
burning; we find that $r_*\simeq 2 R_\odot$ is a typical value for the harmonic mean radius. 
By comparing the number of embedded sources with the number
of T Tauri stars, Kenyon et al. (1990) inferred
a star formation time $\tf\simeq (0.1-0.2)$~Myr
in Taurus-Auriga; to build up a star of average mass ($0.5 \, M_\odot$) in this time
requires an accretion rate $\dot m\simeq (2.5-5)\times 10^{-6}\, M_\odot$~yr\e.
On average, the mass of a protostar will be about half its final stellar mass,
which implies that the typical luminosity of a protostar that
will become a star of average mass is $L\sim (10-20) L_\odot$. 
(The median luminosity of the average mass star is comparable to the median luminosity---see \S \ref{sec:plf} below.)
The median luminosity in the Kenyon et al.~sample is $1.6 L_\odot$.
One statement of the luminosity problem is that the median protostellar luminosity
is observed to be about an order of magnitude less than the expected value.

Kenyon et al.~(1990) also provided an alternative description of the luminosity
problem: Identifying
the peak in the observed luminosity function of embedded
sources at $0.3\, L_\odot$ as the luminosity
of the lowest mass stars, which they took to be $0.1\, M_\odot$ (keep in mind
that this was prior to the discovery of brown dwarfs), and estimating the
radius of these protostars as $r_*\sim 1\, R_\odot$, they inferred a mass accretion
rate of only $10^{-7} \,M_\odot$~yr\e. However, very general arguments 
(Stahler, Shu, \& Taam 1980) indicate
that the accretion rate due to gravitational collapse should be of order
\beqa
\dot m&\sim& \frac{(c_s^2+v_{\rm A}^2+\sigma_{\rm turb}^2)^{3/2}}{G},\\
&\geq &\frac{c_s^3}{G}=1.4\times 10^{-6}\left(\frac{T}{10\,{\rm K}}\right)^{3/2}~~~M_\odot~\mbox{yr\e}.
\label{eq:mdotis}
\eeqa
There are solutions that give higher accretion rates than this, such as
the Larson (1969)-Penston (1969)  solution for the collapse of an isothermal
sphere of constant density, but there are no solutions that give lower 
accretion rates, at least prior to the time that the accretion is affected by the outer
boundary (Henriksen et al 1997).
Since the observed temperature in molecular clouds is $T\sim 10$~K, this
again leads to an order of magnitude discrepancy between observation and theory.

Results from the recent {\it Spitzer} c2d survey of five nearby star-forming
molecular clouds (Evans et al. 2009; the survey does not include Taurus-Auriga) 
confirm that protostars have
low luminosities. Evans et al. (2009) classified the young stellar objects (YSOs)
in the traditional class system, in which Class 0 represents protostars with
envelope masses greater than the protostellar mass; Class I represents
embedded objects with masses greater than the envelope mass; Flat Spectrum
objects represent a transitional class; and Class II corresponds to T Tauri stars.
Dunham et al (2010) analyzed 
this sample and found an 
extinction-corrected median luminosity of $1.5\;L_\odot$.
Including 350~$\mu$m data for many of the brighter sources, they obtained
an extinction-corrected mean 
of $5.3\;L_\odot$.

The luminosity problem is thus well established observationally. It
is fundamental because observations appear difficult to reconcile with some
of the the basic premises of star formation theory, that stars
form via gravitational collapse (eq. \ref{eq:mdotis}) and that they radiate the binding
energy in the process (eq. \ref{eq:lacc}). As the discussion above illustrates,
there are two related aspects to the problem: the observed luminosity appears
to be less than that theoretically expected, and there
is an excess of very low-luminosity sources.

\section{Proposed Solutions}

Kenyon et al. (1990) proposed a number of solutions to the luminosity problem,
of which the two major ones are slow accretion and episodic accretion
(discussed below).
They also suggested that brown dwarfs could alleviate the luminosity
problem by providing sources with luminosities less than they considered.
Since their work predated the discovery of brown dwarfs, they did not
include them. It is now known that brown dwarfs constitute about 20\%
of stars 
(Andersen et al. 2008),
and they do 
permit some sources to have lower luminosities. We include
brown dwarfs in the models described below.

There is an additional effect that reduces the luminosity problem:
The hydromagnetic outflows observed from protostars extract
kinetic energy from the accreting gas (McKee \& Ostriker 2007),
reducing the energy radiated. If half the energy lost by the disk is
mechanical energy and this in turn is half the total potential energy, 
then $f_\acc\simeq \frac 34$ (Offner \& McKee 2010).

\subsection{Episodic Accretion}

Kenyon et al. (1990)
suggested that the accretion onto the protostar (as opposed to infall
onto a circumstellar disk) could be episodic, so that much of the protostellar mass would be accreted in short periods of time, with high
luminosities. Such brief, high-luminosity accretion events could be associated
with FU Ori outbursts, which have inferred accretion rates of $\dot m\sim 10^{-4}\,
M_\odot$~yr\e. Kenyon et al. (1990) pointed out that the 150 YSOs in
Taurus Auriga that formed in the last $10^6$~yr correspond to a total
accretion rate of $0.75\times 10^{-4}\,M_\odot$~yr\e, comparable to that of a single FU Ori
object; thus, a signficant fraction of the mass of a protostar might be
acquired during FU Ori events.

Hartmann \& Kenyon (1996) refined this argument: They cited a
rate of star formation within 1~kpc of the Sun of $(5-10)\times 10^{-3}\,M_\odot
$~yr\e, which is similar to the recent estimate of $8\times 10^{-3}\, M_\odot$~
yr\e\ by Fuchs et al (2009). At that time there were 5-9 known FU Ori objects
within 1~kpc; if each were to accrete at a rate of $10^{-4}\, M_\odot$~yr\e, then
protostars could gain about 10\% of their mass this way. They pointed out that
this was a lower limit, since more FU Ori objects would most likely be discovered.
That has indeed occurred:
Greene et al. (2008) count a total of 22 known FU Ori objects, of which 18 are within
1 kpc. Rounding off, we infer that
20 objects accreting at $10^{-4}\, M_\odot$~yr\e\ would account for
25\% of the mass accreted by protostars within 1 kpc. 

To elaborate on this result, consider a simple model for
episodic accretion, in which protostars are in
a high-accretion state for a total time $\dth$
and in a low-accretion state for the rest of the time, 
which is effectively the star formation time, $\Delta t_\low\simeq
\tf$. The fraction of the mass accreted during a high state is
\beq
F_\high=\frac{ \mdh\dth}{\avg{\mf}}.
\eeq
We shall adopt parameters appropriate for FU Ori outbursts, but in fact all
episodes of high accretion could be included in $F_\high$. The total
time spent in the high state is
\beq
\dth=\frac{\caln_{p,\,\high}}{\dot\caln_*},
\label{eq:dth}
\eeq
where $\caln_{p,\,\high}$ is the number of protostars in a high state in some volume of
the Galaxy and $\dot\caln_*$ is the star formation rate there.
For a mean stellar mass of $0.5\, M_\odot$, the local star formation rate cited above
corresponds to a birthrate within 1 kpc of
the Sun of 0.016 stars yr\e.
If the number of FU Ori objects within 1 kpc is about 20, then
$\dth=1250$~yr. Inserting Hartmann \& Kenyon's estimate 
that $\mdh\simeq 10^{-4}\,M_\odot$~yr\e, we recover the result given above,
$F_\high\simeq 0.25$. We also note that the duty cycle of FU Ori outbursts is
small: $\Delta t_\high/\tf\sim 10^3/(10^{5-6}) <0.01$. This small value is
consistent with the absence of any known FU Ori sources in the Evans et al.~(2009) sample. 

There are two uncertain numbers that entered into this estimate of $F_\high$,
the number of FU Ori objects, $\caln_{p,\,\high}$, and the accretion rate,
$\mdh$. Undoubtedly, more
such objects will be discovered within 1 kpc of the Sun in the future; however,
it should be noted that several of the 
bursting objects, such as L1551 IRS5, have luminosities,
and therefore accretion rates,
an order of magnitude less than the average
(Hartmann \& Kenyon 1996). 
Further study will also refine the observed
value of the mean accretion rate. Hartmann \& Kenyon
(1996) infer an accretion rate $\mdh=1.9\times 10^{-4}\, M_\odot$~yr\e\
and a stellar radius $r_*=5.9\, R_\odot$ for FU Ori itself. However, after
the rapid accretion ceases, the star will shrink back to its original size,
releasing a comparable amount of energy. A mean value $\mdh\simeq 1\times 10^{-4}\, M_\odot$~yr\e\ thus seems reasonable in this case.

Hartmann \& Kenyon (1996) also 
noted that given 5 known outbursts in 60 years and a star formation rate of
0.01 stars yr\e\ within 1 kpc of the Sun implies that there are about
10 bursts per object. The more recent data cited above
lead to a similar conclusion. Since $\dth\simeq 1250$~yr, this means that
a typical outburst lasts about 100 yr, consistent with the
observationally inferred lifetime (Hartmann \& Kenyon 1996).

Finally, we note that a potential problem with the
FU-outburst solution to the luminosity problem is that
these outbursts appear to be located preferentially in regions of low 
star-formation rates (Greene et al 2008).

\subsection{Slow Accretion ($\dot m\la 10^{-6} M_\odot$ yr\e)}

As an alternative solution to the luminosity problem,
Kenyon et al. (1990) suggested two ways to increase the
ages of the protostars and therefore reduce the inferred accretion rate:
(1) If the star formation rate decreased substantially over the  lifetime of the
T Tauri stars, then the observed number of Class I sources would require
a greater lifetime; or (2) if the lifetime of the T Tauri stars were larger, then
the protostellar lifetime would increase proportionately. 
They rejected the first possibility since it is difficult to understand how
star formation can decelerate in 1~Myr when it is in a region with a crossing
time of 10~Myr. The second possibility is more plausible, since the estimated lifetime of T Tauri stars has increased in the intervening 20 years. 
In their
recent analysis of protostellar lifetimes, Evans et al (2009) concluded that the lifetime
of Class II sources is about 2 Myr, which would imply a Class I lifetime
of $\tf\simeq (0.2-0.4)$~Myr in Taurus Auriga. The larger of these two values
is favored by data
in the c2d survey, for which Evans et al.~(2009)
found a Class I lifetime of 0.44~Myr.
The total protostellar lifetime, $\tf$, must also include the time spent in the Class 0 phase,
which they found to be $\simeq 0.1$~Myr, so that the total inferred mean lifetime
of protostars in these clouds is $\avg{\tf}=0.54$~Myr. 
This long protostellar lifetime, together with the correction for non-radiative
energy losses ($f\_{\rm acc}=\frac 34$) and the correction for unseen outbursts
(also a factor of $\frac 34$) 
effectively resolves the luminosity problem: The accretion rate for a star of average mass is then $\dot m\simeq 10^{-6}\, M_\odot$~yr\e,
which corresponds to a typical accretion luminosity from equation (\ref{eq:lacc}) of
about $2.2\, L_\odot$, comparable to the observed median luminosity
of $1.5\, L_\odot$. 

What about the other aspect of the luminosity problem,
which is that the inferred accretion rates 
are much less than expected theoretically? There are two physical
effects that reduce the accretion rate below that in standard models, namely,
protostellar outflows and the finite size of the protostellar envelope.
Bontemps et al.~(1996)
found that protostellar outflow rates vary as the envelope mass and
showed that this could be understood if accretion rates declined exponentially
with time:  $\dot m=\dot m_0\exp(-t/t_*)$ (see also Myers et
al.~1998), where
the decay time is $t_*=\mf/\dot m_0$ and $\mf$ is the final protostellar
mass (ie., the initial stellar mass). 
This exponential decline in the accretion rate does not capture the
reduction due to the initial stage of protostellar outflows (Shibata \& Uchida 1985).
It is thus plausible when these effects are included, current theories can be
consistent with the inferred low accretion rates. However, this will
be possible only if the high accretion rates associated with an initial
Larson-Penston accretion phase (Henriksen et al. 1997, Schmeja \& Klessen 2004)
do not release a signficant amount
of radiative energy.

\section{The Protostellar Luminosity Function}
\label{sec:plf}

\subsection{Previous Work}

Any solution of the luminosity problem must explain
the distribution of protostellar luminosities--the protostellar luminosity
function (PLF). Kenyon et al.~(1990) introduced the PLF and
considered two cases: (1) the standard isothermal sphere (IS) case
(Shu 1977), in which the accretion rate is independent of mass
so that the formation time, $\tf$, is proportional to the final mass, $\mf$;
and (2) a case in which the accretion rate is proportional to 
$\mf$, so that $\tf$ is independent of $\mf$. Fletcher and Stahler
(1994a,b) extended this work in the IS case by determining both the protostellar mass
function (PMF) and the PLF for pre-main-sequence stars. 

Dunham et al.~(2010) carried out a detailed analysis of the joint distribution of
protostellar luminosities and bolometric temperatures. Interestingly,
they adopted an accretion rate of $4.6\times 10^{-6}\, M_\odot$~yr\e,
much higher than implied by the protostellar lifetimes inferred by
Evans et al. (2009). In one series of models, they included outflows,
which entrained envelope material and resulted in a core star-formation
efficiency of 0.3-0.5. The outflows were not intrinsically collimated,
as in the study by Matzner \& McKee (2000), so that outflow cavities eventually
expanded to an opening angle of $90 \deg$, terminating accretion.
The outflows led to substantial variations in the observed bolometric luminosity
and temperature as functions of the inclination angle.
Despite the reduction in the accretion rate, the mean luminosities were still
too high.
The best agreement with observation was obtained by assuming
that most of the mass was accreted in FU Ori-type events with accretion rates of $10^{-4}\,
M_\odot$~yr\e, although the resulting models spent much more time with
bolometric temperatures above $10^3$~K than is observed. These models
required $F_\high\sim 0.8$ (Dunham, private communication). This is
consistent with observation only if the number of FU Ori sources
within 1 kpc of the Sun were about 3 times higher than currently known (see eq.
\ref{eq:dth}).

\subsection{The PMF and the PLF}

We follow McKee \& Offner (2010) and Offner \& McKee (2010) in
describing the protostellar mass function (PMF) and the protostellar luminosity
function (PLF). Let
$\psi(\mf)d\ln \mf$ be the fraction of stars born in the mass range $d\mf$;
$\psi$ is thus the IMF. The PMF is the present-day mass function of
the protostars, and it must be consistent with the IMF when the
protostellar mass reaches its final value, $\mf$. 
We denote the fraction of protostars in the mass range $dm$ that will
have final masses in the mass range $d\mf$ by $\ppt(m,\mf)d\ln m d\ln\mf$;
it is the IMF weighted
by the time the protostar spends at a given mass,
\beq
\ppt(m,\mf)=\left[\frac{(m/\dot m(m, \mf))}{\avg{\tf}}\right] \psi(\mf),
\label{eq:ppt}
\eeq
where $\avg{\tf}$ is the average protostellar lifetime.
We have assumed that the accretion rate is a function of $m$ and $\mf$,
which can be readily generalized to allow for
high and low accretion states.
The PMF is the integral of this function over all possible values of
the final mass, ranging from the lower bound on the IMF, $m_\ell$,  
to the upper bound, $m_u$, subject to the constraint 
$m \leq \mf$:
\beq
\psi_p(m)=\int_{\max(m_\ell, m)}^\mup \ppt \;d\ln\mf.
\eeq

It follows that the fraction of protostars in the mass range $dm$ and the luminosity
range $dL$ is
\beq
\Ppt(L,m)\,d\ln m\,d\ln L=\ppt(m,\mf)\, d\ln m\, d\ln\mf.
\eeq
The PLF is then
\beqa
\Ppl&=&\int_{m_{\rm min}}^{m_{\rm max}}  d\ln m\Ppt(L,m),\\
&=&\int_{m_{\rm min}}^{m_{\rm max}} d\ln m \;\frac{\ppt[m,\mf(L,m)]}{|\partial \ln L/\partial\ln\mf|},
\eeqa
where the limits of integration are such that $\ml \leq \mf \leq m_u$ 
and $m\leq\mf$.
The PMF and PLF depend on the history of the
mass accretion rate, $\dot m(m,m_f)$ (eq. \ref{eq:ppt}) \textit{and therefore on
the theory of star formation.}

\section{Testing Theories of Star Formation}

\subsection{Accretion Histories}

We consider three different theories of star formation, plus one variant:
\begin{itemize}
\item[*] {\bf Isothermal Sphere} (IS, Shu 1977)---Inside-out collapse of singular isothermal sphere:
\beq
\dot m = \dot m_{\rm IS}=1.54\times 10^{-6} (T/10\,{\rm K})^{3/2}~~~M_\odot~\mbox{yr\e}.
\eeq
The formation time is proportional to the mass, $t_f\propto m$, so this accretion model is valid only for low-mass stars (Shu, Adams \& Lizano 1987).
\bigskip
\item[*] {\bf Turbulent Core} (TC, McKee \& Tan 2002, 2003)---Inside-out collapse of a turbulent core:
\beq
\dot m =\dot m_{\rm TC} \left(\frac{m}{m_f}\right)^{1/2} m_f^{3/4},~~~\mbox{with\ }
\dot m_{\rm TC}\propto \Sigma^{3/4}.
\eeq
The accretion rate increases with both the protostellar mass, $m$, and the final
mass, $\mf$. This model was developed for high-mass star formation, and is
equivalent to the accretion rate in the Hennebelle \& Chabrier (2008) theory of the IMF.
\bigskip
\item[*] {\bf Two-Component Turbulent Core} (2CTC)---Begins as isothermal accretion and evolves
to turbulent accretion (similar to the TNT model of Myers \& Fuller 1992):
\beq
\dot m = \left[\dot m_{\rm IS}^2 + \dot m_{\rm TC}^2 \left(\frac{m}{m_f}\right) m_f^{3/2}\right]^{1/2}.
\eeq
\bigskip
\item[*] {\bf Competitive Accretion} (CA, Zinnecker 1982, Bonnell et al.
1997)---Stars form in a common gas reservoir, usually accreting at the tidally limited Bondi rate, 
$\dot m \propto m^{2/3}$, and all having the same formation time:
\beq
\dot m=\dot m_{\rm CA}\left(\frac{m}{m_f}\right)^{2/3}m_f,~~~\mbox{with\ } \dot m_{\rm CA}
\propto 1/\mbox{(free-fall time)}.
\eeq
\end{itemize}

\begin{figure}
\vspace*{-0.5 cm}
\includegraphics[width=5.3in]{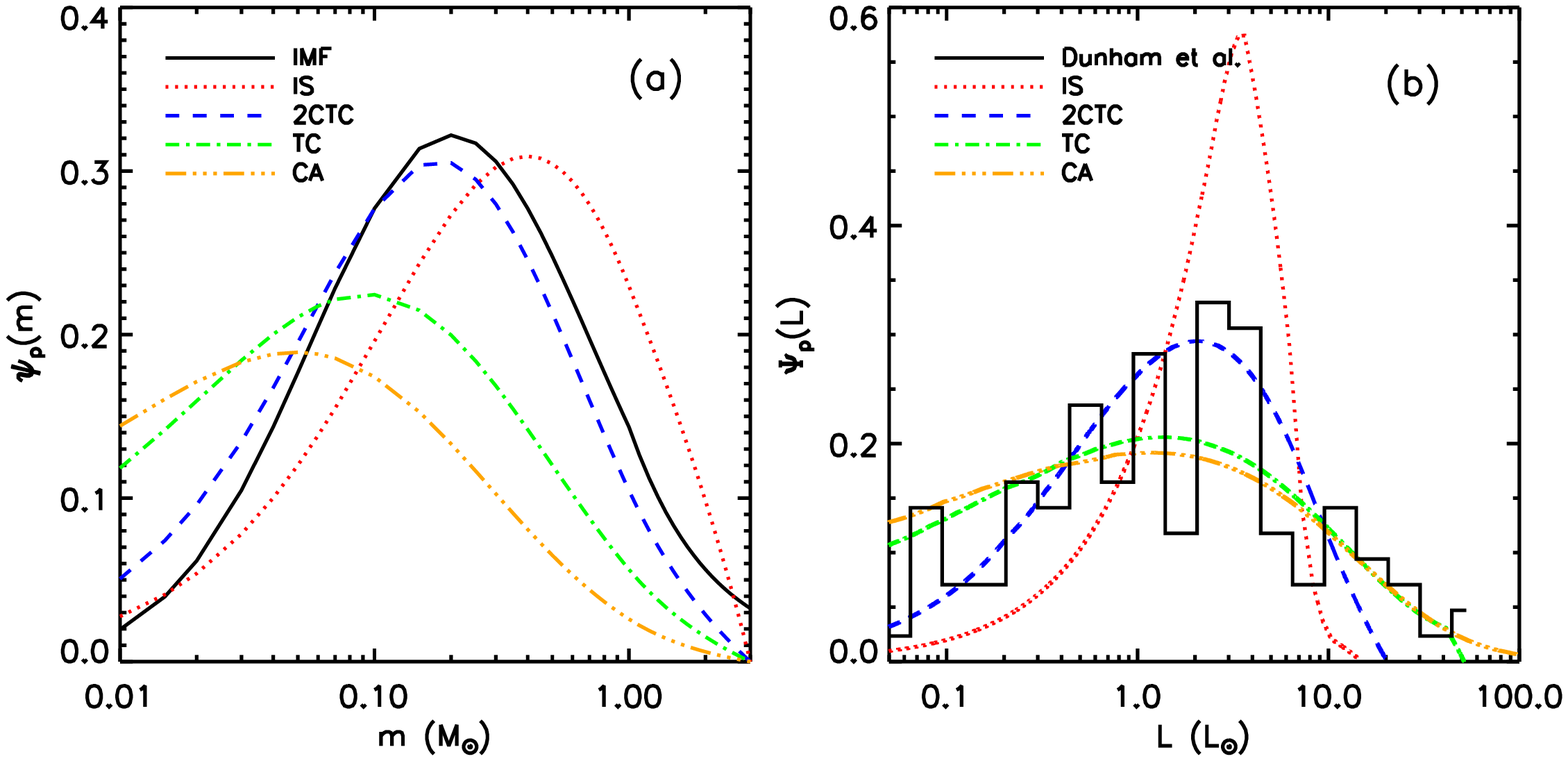}
\vspace*{-0.5 cm}
\caption{Left: The PMF, $\psi_p(m)$, for the four accretion models 
and the Chabrier IMF. Right: The PLF, $\Psi_p(L)$, for the same
models 
assuming 
no tapering of the accretion rate,
with $F_\high = 0.25$ and $\avg{\tf} =0.56$ Myr with the data 
from Dunham et  al.~(2010). Both panels adopt a cluster upper mass $m_u =
3~M_\odot$.
\label{pmfplf}}
\end{figure}

\subsection{Comparing the PMF and PLF with Observation}

We adopt a truncated Chabrier (2005) IMF:
\beqa
\psi(\mf)&\propto& \exp-\left[\frac{\log^2(\mf/0.2)}{2\times 0.55^2}\right]~~~~
(\mf\leq 1\; M_\odot)\\
&\propto& \mf^{-1.35}~~~~~~~~~~~~~~~~~~~~(1\; M_\odot <\mf\leq m_u)
\eeqa
The Evans et al.~(2009) sample of Class II YSOs has about 400 objects.
In this sample, 9 YSOs are expected with masses exceeding $3 M_\odot$
if $m_u\gg 3 M_\odot$ (i.e., stars with masses much greater than $3M_\odot$
are possible).
Since none are seen, we adopt an upper cutoff  $m_u=3 M_\odot$ for
the IMF.

Figure \ref{pmfplf}a illustrates the PMF (eq. 3.2) estimated with each of the
accretion histories. 
As discussed by McKee \& Offner (2010), the
models predict very different mass distributions. For example, in the
isothermal sphere
case more massive protostars spend a
longer time accreting and consequently weight the distribution towards higher
masses. In contrast, for competitive accretion, where all the protostars share the same
protostellar lifetime,  the significantly larger number
of low-mass protostars shifts the PMF towards lower masses.
Unfortunately, since we can't directly measure the protostellar masses,
Figure \ref{pmfplf}a is not sufficient to observationally 
discriminate between the models.  

For a more direct comparison, Offner \& McKee (2010) calculate the
PLF using the predicted mass functions in combination with a stellar
evolution model (see Offner et
al.~2009 for details). Figure \ref{pmfplf}b shows the PLF for
each accretion model plotted with the extinction corrected luminosities from
Dunham et al.~(2010). The model curves assume that
25\% of the mass is accreted during unseen bursts ($F_\high = 0.25$)
and $\avg{\tf} = 0.56$ Myr. This is equivalent to applying both the slow
accretion and variable accretion solutions and thus likely represents a
lower bound on the predicted luminosities. For comparison with Dunham et
al. 2010, we adopt an upper bolometric luminosity uncertainty of 50\% and use the uncorrected
bolometric luminosities to set a lower error bound.

The distributions can be
characterized by the mean, median, and standard deviation. 
For $m_u = 3~M_\odot$ and $L_{\rm min}= 0.05~L_\odot$, the fiducial 
models have means in the range 2.5 $L_\odot$ (2CTC) - 3.6 $L_\odot$ (CA), all
a factor of $\sim$1.5-2 below the observed value: $5.3~\frac{+2.6}{-1.9}~L_\odot$.
We find that the mean luminosities fall in a narrower range, 2.6 $L_\odot$
(2CTC)- 3.4 $L_\odot$ (IS), when the accretion rates taper off towards the end of the protostellar lifetime:
\beq
\dot m = \dot m_0(m, \mf) \left[1-\left({{t}\over{\tf}}\right) \right],
\eeq
where $\dot m_0 (m,\mf)$ is the untapered accretion rate.
(Foster \& Chevalier 1993 found that the accretion rate tapered off at late times
in their 1D calculations, and Myers et al.~1998 included an exponential tapering in their models.)
Only the non-tapered CA and tapered IS models fall within the uncertainty,
suggesting that the adopted lifetime may be too high by as much as a
factor of 2.
In contrast, the fiducial medians, which range from 0.8 $L_\odot$ (CA) to 2.5 $L_\odot$
(IS), are in better agreement with the observed median of $1.5~\frac{+0.7}{-0.4}
~L_\odot$. The median of the TC ($0.9~L_\odot$) and 2CTC ($1.4~L_\odot$) models are within error and remain
so even for a lifetime reduced by a factor up to 2.4 and 1.6, respectively.
The observed standard deviation of $\log L$,
$0.7 \frac{+0.2}{-0.1}$, is consistent with the 2CTC (0.6), TC (0.7) and
CA (0.8) cases.
The outcome of the comparison is sensitive to the values of $F_\high$
and $f_{\rm acc}$, in addition to $\avg{t_f}$, and all have
significant uncertainty. The parameter dependence may be reduced by comparing to
the ratio of the median to the mean luminosity, $0.3~\frac{+0.2}{-0.1}$. This rules out the IS model.

\section{Conclusions}

The luminosity problem in low-mass star formation can be resolved if low-mass 
stars form slowly, over a period $\sim 0.5$~Myr, as suggested by
the results of Evans et al. (2009). Indeed, if some of the accretion energy
is released mechanically ($1-f_\acc\simeq 1/4$) and in unseen FU Ori outbursts
($F_\high\simeq 1/4$), then 
for most models
the star formation time must be somewhat
less than 0.5 Myr to be consistent with the observed protostellar luminosities.
Different theories of star formation predict different prototellar mass functions,
which are currently inaccessible to direct observation, and protostellar
luminosity functions, which have been observed.
The latter serves as an important metric to discriminate between
theories of star formation.

\acknowledgements We thank Michael Dunham, Melissa Enoch, and
Neal Evans for discussions of their work, and Philippe Andre, Lee
Hartmann, Jean-Francois Lestrade, and
Mark Krumholz for useful comments.
This research has been supported by the NSF under grants AST-0908553
(CFM) and AST-0901055 (SSRO).
CFM also acknowledges the support of the Groupement d'Int\'er\^et Scientifique
(GIS) ``Physique des deux infinis (P2I)."

\newcommand{\apj}{\textit{ApJ}}
\newcommand{\apjl}{\textit{ApJL}}
\newcommand{\aap}{\textit{A\&A}}
\newcommand{\araa}{\textit{ARAA}}
\newcommand{\mnras}{\textit{MNRAS}}

\end{document}